\documentclass[usenatbib,useAMS]{mn2e}
\usepackage[table]{xcolor}
\usepackage[dvips]{graphicx}
\usepackage{datetime}
\usepackage{dcolumn}
\usepackage{threeparttable}
\usepackage{multirow}
\usepackage{multirow}
\usepackage{paralist}

\newcommand{\aap}{A\&A} 
\newcommand{\aj}{AJ} 
\newcommand{\apj}{ApJ} 
\newcommand{\mnras}{MNRAS} 
\newcommand{\apjl}{ApJ} 
\newcommand{\pasa}{PASA} 

\newcommand{\ozqtwo}{1$\to$0Q(2)~} 
\newcommand{\ozqo}{1$\to$0Q(1)~} 
\newcommand{\ozqt}{1$\to$0Q(3)~} 
\newcommand{\ozso}{1$\to$0S(1)~} 
\newcommand{\ozst}{1$\to$0S(3)~} 
\newcommand{\ozstwo}{1$\to$0S(2)~} 
\newcommand{\oozso}{1$\to$0S(1)} 
\newcommand{\ozsoo}{1$\to$0S(0)} 
\newcommand{\tost}{2$\to$1S(3)} 
\newcommand{\toso}{2$\to$1S(1)~} 
\newcommand{\halpha}{H$\alpha$}
\newcommand{\kms}{km~s$^{-1}$}
\newcommand{\simm}{$\sim$}
\newcommand{\hco}{HCO$^{+}$}

\newcommand{\asin}{sin$^{-1}$}
\newcommand{\Rmin}{$R_{\rm min}$}
\newcommand{\Rmax}{$R_{\rm max}$}
\newcommand{\rto}{1$\to$0/2$\to$1S(1)}
\newcommand{\kband}{$K$-band}
\newcommand{\ohtto}{OH~231.8+4.2}

\title[Shocked H$_2$ around OH~231.8+4.2]{Discovery of shocked H$_2$ around OH~231.8+4.2}
\author[K.P. Forde, and T.M. Gledhill]{K.P. Forde$^{1}$\thanks{E-mail:
k.p.forde@herts.ac.uk}, T.M. Gledhill$^{1}$\\
$^{1}$Centre for Astrophysics Research, University of Hertfordshire, College
Lane, Hatfield, AL10 9AB, UK}

\begin{document}

\date{Accepted 2011 Month ??. Received 2011 Month ??; in original form 2011 Month ??}

\pagerange{\pageref{firstpage}--\pageref{lastpage}} \pubyear{2009}

\maketitle

\label{firstpage}

\begin{abstract}

We present $K$-band integral field observations of the 
circumstellar envelope of the evolved star \ohtto. Spatial and spectral
information were simultaneously acquired using
the {\sc sinfoni} integral field unit, with adaptive optics, on the Very Large Telescope. 
The observations reveal the discovery of H$_2$ emission (1) around the
centre of the nebula and (2) located in clumps along the Western side of the
Northern lobe, 
presumably associated with the strong shocks that stimulate the previously
reported \halpha~emission at the same location. 
An observed H$_2$ \rto~line ratio of ~8.3$\pm$1.9 was calculated for the central
field, a value consistent with shock excitation.

\end{abstract}


\section{Introduction}
\label{sec:Intro}

OH~231.8+4.2 (hereafter OH231) is an O-rich late spectral
type (M) central star (Mira variable, QX Pup) with bipolar high-velocity
dust and gas outflows~\citep{2001A&A...373..932A},  filamentary structures observed in
scattered and molecular line emission, and large angular size
(10$^{\prime\prime}$$\times$60$^{\prime\prime}$). 
Often labeled as a post-AGB object or pre-planetary nebula, the
presence of both a Mira central star and a main-sequence companion of spectral
type A \citep{2004ApJ...616..519S} seems to contradict this classification. OH231
is more likely a D-type bipolar symbiotic system
\citep[][]{2010PASA...27..129F}. However in some cases, morphological similarities do exist
between post-AGB and symbiotic objects, most strikingly the presence of
highly collimated bipolar nebulae. It is via fast-collimated outflows that these
stars shape their surrounding nebula. 
Understanding the development and origin of these fast outflows
is critical for advancing
hydrodynamical models of wind interaction. Recent work by
\citet{2009ApJ...696.1630L} attempting to
reproduce the high velocity molecular emission in AFGL~618 using 
collimated fast wind models, emphasises the need for further observations and
model development in this area.

OH231 has been the
subject of many studies spanning multiple wavelength ranges, for example:
\citet{1985ApJ...297..702C} were first to propose the existence of a binary companion;
\citet{2002A&A...389..271B} imaged the
shape of the shocks using H${\alpha}$ (reproduced in
Fig.~\ref{fig:oh231_naco_sin_wfpc2} [a]) detected with the {\em Hubble Space
Telescope (HST)}; \citet{2003ApJ...585..482M} report {\em HST}/NICMOS NIR
images of the dust distribution and hence a high resolution map of the
extinction through the nebula. \citet{2006ApJ...646L.123M} using the MIDI
and NACO instruments on the Very
Large Telescope (VLT) detected a compact circumstellar disc. The envelope of
OH231 is also known to be rich in molecular species (e.g. H$_2$O, OH,
 and SiO) 
however previous studies in the NIR have all returned null detections of H$_2$
\citep[e.g.][]{1998ApJ...509..728W,2006ApJ...646L.123M}.

In this Letter, we present the results of preliminary observations of OH231 at
NIR ($K$-band) wavelengths showing for the first time the presence of shock-excited H$_2$. 
Throughout this work we assume OH231 is a member of the open cluster M46 at a
distance of 1.3~kpc~\citep{1985ApJ...292..487J}. The origin of the
coordinate system used in Figures~1,3, and 4 is given by the location of the SiO maser emission at
RA=07$^h$42$^m$16$^s$.93,
Dec=-14$^\circ$42$^{\prime}$50$^{\prime\prime}$.2 (J2000)~\citep{2002A&A...385L...1S},
and the inclination angle of the bipolar axis is 36$^\circ$~to the plane
of the sky \citep{1992ApJ...398..552K}.

\section{Observations and Data Reduction}
\label{sec:obs}

The data were taken using the {\sc sinfoni}~\citep{2003SPIE.4841.1548E}
instrument located on UT4 at VLT at Paranal, Chile, on the 1$^{\rm st}$/2$^{\rm nd}$ Feb 2010. 
We use the lowest resolution mode (LRM), corresponding to the widest
field-of-view (8$^{\prime\prime}$ x 8$^{\prime\prime}$) with adaptive optics
(AO) and a natural guide star (NGS). A plate scale
of 250$\times$125 mas pixel$^{-1}$, and a spectral and velocity resolution of
4580 and 66 km~s$^{-1}$, respectively, are available at this resolution
(for a dispersion of 2.45~\AA/pix and line FWHM of 1.96 pixels).   
All observations utilised the $K$-band (2.2~$\umu$m) filter which covers many
ro-vibrational H$_2$ emission lines. 
The ambient seeing varied from $\sim$~0.6$^{\prime\prime}$ to 1.1$^{\prime\prime}$ during the observations. 
The OH231 observations consisted of three fields focused on (1) the edge of 
the Northern lobe, (2) the central region, and (3) the middle of the Southern
lobe (labelled N, C, S in Fig.~\ref{fig:oh231_naco_sin_wfpc2} [b]).
No H$_2$ was detected in the Southern field and will not be discussed further. 
Table~\ref{tab:sum_obs} summarises exposures for
each of the three fields. Telluric
standard stars used for calibration are HD~75004 (G0V), and HD~63487 (G2V) for
night one and two, respectively. An average AO-corrected PSF of \simm 340 mas
FWHM is estimated from the standard stars. The data were reduced using the ESO common pipeline library to a
wavelength-calibrated datacube and further analysed using both PyRAF\footnote{PyRAF is
a product of the Space Telescope Science Institute, which is operated by AURA
for NASA.}.
Wavelength calibration errors were corrected by comparison of OH emission lines
with a high resolution template. Quoted velocities were adjusted to local standard of
rest (LSR) velocities using 22.88 and 23.30~\kms~corrections for night one and
night two, respectively.
Line maps (Fig.~\ref{fig:oh231_naco_sin_wfpc2} [c,d]) were generated by
fitting the H$_2$ emission lines with a Gaussian profile; initial fit parameters
(FWHM$_0$, central wavelength$_0$, etc.) were determined by manually fitting an
individual H$_2$ line, only lines with a FWHM $\approx$ FWHM$_0$ make up the final
line maps. Signal-to-Noise (S/N) of the line maps was enhanced by smoothing the
data with a 2$\times$2 pixel boxcar filter. 
Line rest wavelength information is from~\citet{1987ApJ...322..412B}. 

\begin{table}
\caption{Details of the VLT/{\sc sinfoni} observations for OH231 and
the telluric standard stars taken on 1$^{\rm st}$/ 2$^{\rm nd}$ February 2010.}
\begin{threeparttable}
\begin{tabular}{@{}l@{}cc@{}cc@{}ccc}
\hline
\hline
			&\multicolumn{4}{c}{OH231} & &\multicolumn{2}{c}{HD}\\	\cline{2-5}\cline{7-8}\noalign{\smallskip}
			&	South &\multicolumn{2}{c}{Central}     & North&&75004	 & 63487 	\\
\hline                                                                                     
Tot. Exp. (secs)    &	600	&	600	&	480	&   800 &&   1	&   10 \\
K (mag)		& 	-- 	     &	--	&  --	&     --     && 7.268$^{\dagger}$	&  7.674$^{\dagger}$	\\
Night			& 	1 	     &	1	&  2	&     1     &&	1	&  2	\\
\hline
\end{tabular}
\label{tab:sum_obs}
\begin{tablenotes}
\item[$\dagger$] 2mass magnitude from: SIMBAD
\end{tablenotes}
\label{tab:comptab}
\end{threeparttable}
\end{table}

\section{Results}
\label{sec:h2emis}

\begin{figure*}
\begin{center}
\includegraphics[width=0.87\textwidth,height=0.6\textwidth,angle=0]{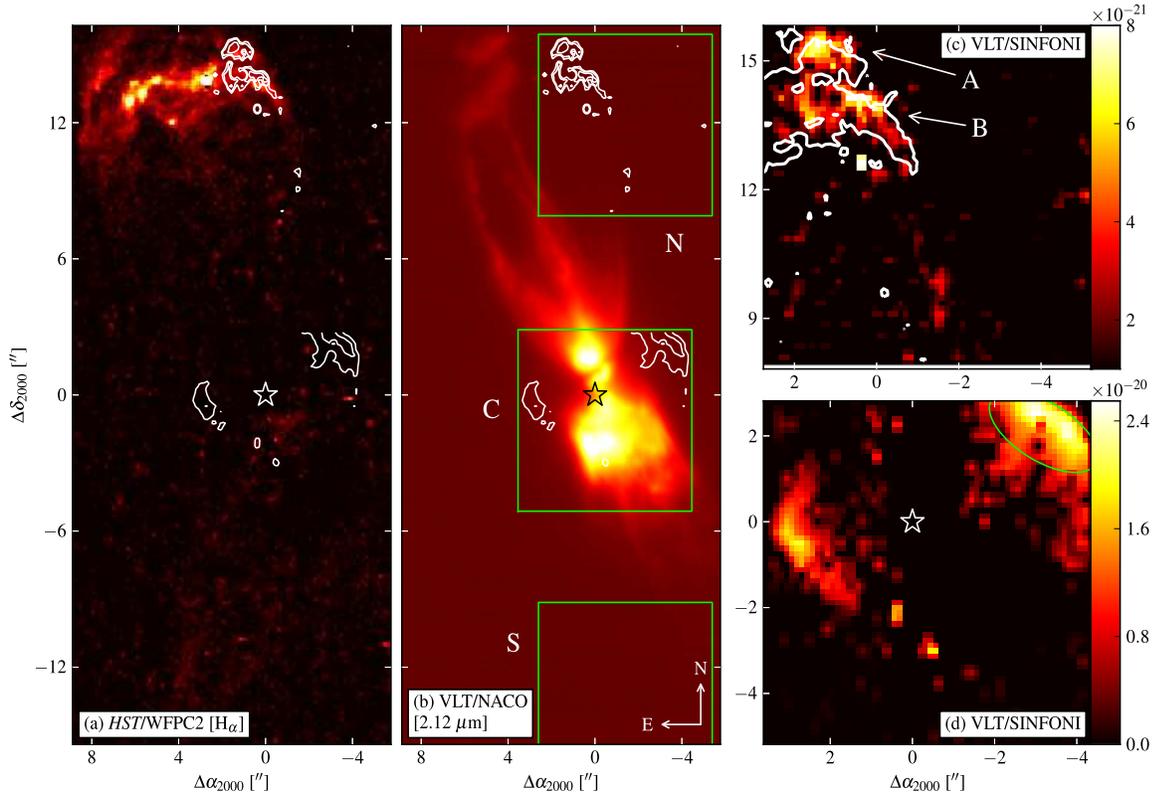}
\caption{Narrow-band images and line maps for OH231; a star symbol indicates the
SiO maser position (see text), inset labels display the
telescope/instrument information. $(a)$ Continuum-subtracted~\halpha~image
with the \ozso H$_2$ contours overlaid in white, $(b)$ 2.12 $\umu$m image,
showing mainly scattered light,
with H$_2$ contours overlaid, {\sc sinfoni} North, Central, and South (labeled N, C, S) fields are marked by green squares, inset compass shows orientation, 
$(c)$ \ozso H$_2$ line map showing the extent of the emission located 
along the edge of the Northern lobe, two knots (A, B) of H$_2$
emission are indicated by arrows; superposed contours indicate the location of
the \halpha~emission,
$(d)$ \ozso H$_2$ line map for the central region showing the
position/extent of the region used for \rto~calculation (ellipse). Colour bar units are in W m$^{-2}$ px$^{-1}$. 
The \halpha~data, from the Hubble Legacy Archive (see
Acknowledgments), has been continuum-subtracted using the procedure outlined in
\citet{2002A&A...389..271B}. The 2.12 $\umu$m data was first published in
\citet{2006ApJ...646L.123M} and has been reprocessed using the ESO pipeline.}
\label{fig:oh231_naco_sin_wfpc2}
\end{center}
\end{figure*}

\begin{figure}
\includegraphics[width=0.5\textwidth,height=0.3\textwidth,angle=0]{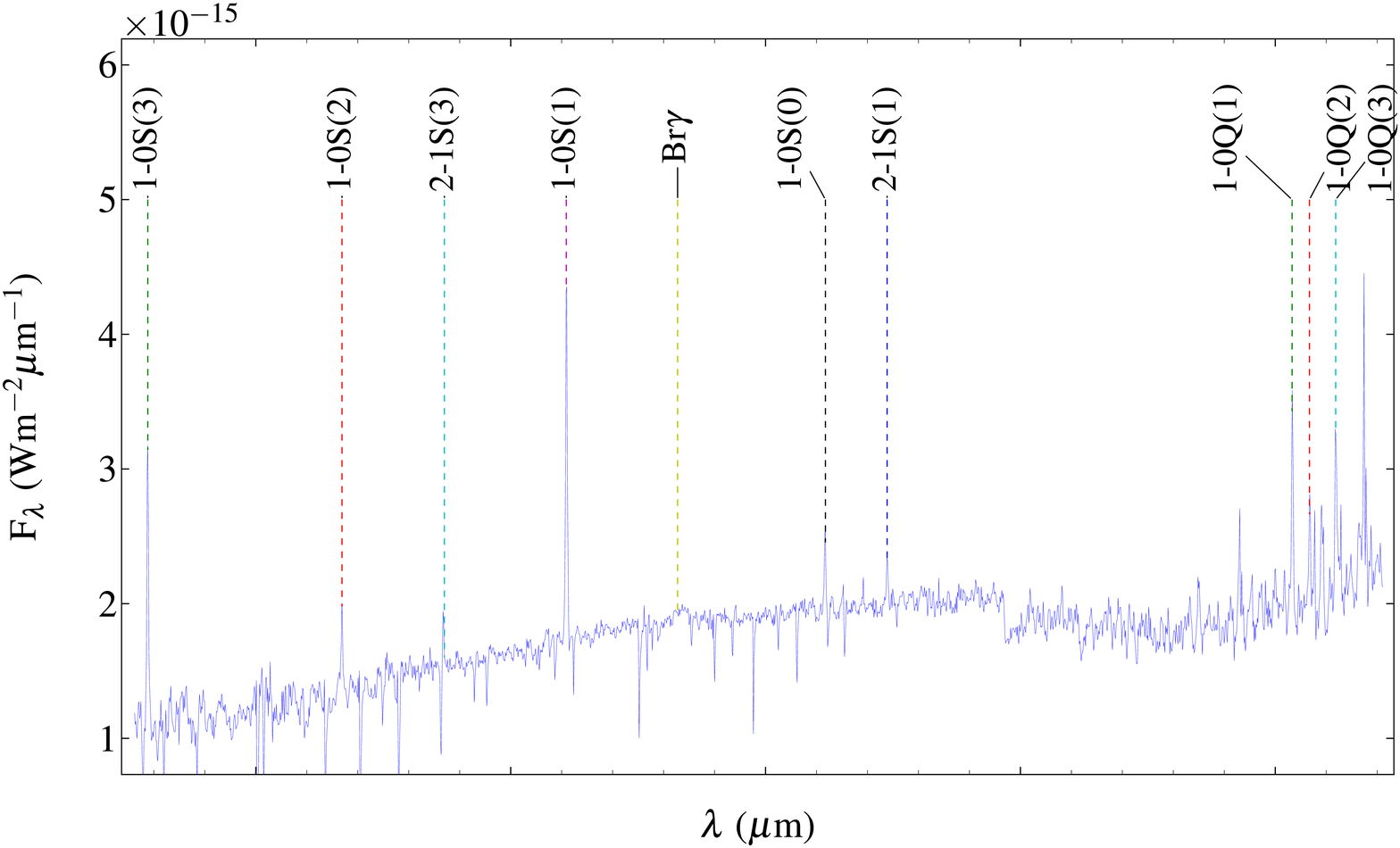}
\caption{An integrated spectrum of the Western side of
the central H$_2$ showing the detected lines, the position of the Br$\gamma$
recombination line is shown for reference. Extracted spectra location marked in
Fig.~\ref{fig:oh231_naco_sin_wfpc2} (d).}
\label{fig:oh231_spec}
\end{figure}

We report the detection of several H$_2$ emission lines arising from both
the centre and Northern lobe of OH231. 
In Figure~\ref{fig:oh231_naco_sin_wfpc2} (c,d) we present continuum-subtracted
line maps of the \oozso~
transition for both fields, showing clearly H$_2$ emission
arising from the centre of OH231 (Fig.~\ref{fig:oh231_naco_sin_wfpc2} [d]) and
from knots of material in the Northern
lobe (Fig.~\ref{fig:oh231_naco_sin_wfpc2} [c]). 
In Fig.~\ref{fig:oh231_naco_sin_wfpc2} (a) the
contours of the detected H$_2$ are shown in relation to \halpha~emission, while
Fig.~\ref{fig:oh231_naco_sin_wfpc2} (b) 
gives the location of the H$_2$ relative to the strong continuum emission as
imaged with a 2.12 $\umu$m filter. 
An integrated spectrum of the Western region of the centrally located H$_2$ is shown in
Figure~\ref{fig:oh231_spec}, from which we note, 
\begin{inparaenum}[a)]
\item{several S- and Q-branch ro-vibrational H$_2$ lines, and }
\item{a CO bandhead absorption feature (\simm 2.3 $\umu$m).}
\end{inparaenum}

Channel maps, extracted from the {\sc sinfoni} datacube, are presented in Figures~\ref{fig:oh231_ch_maps_cen} and
\ref{fig:oh231_ch_maps} for the central and Northern regions respectively, showing how the distribution of the
H$_2$ changes across the line profile. 

\begin{table}
\caption{Line fluxes, $F$, (in units of 10$^{-19}$ W m$^{-2}$) of the observed
H$_2$ lines for the central and
Northern fields. Measured peak line wavelengths (in $\umu$m) are given for the central field.}
\begin{threeparttable}
\begin{tabular}{lll@{ $\pm$ }llll}
\hline
\hline
			&	\multicolumn{2}{c}{Central} &&&& \multicolumn{1}{c}{Northern} \\ \cline{2-4}\cline{6-7}
\multicolumn{1}{c}{Line}&\multicolumn{1}{c}{$\lambda_{\rm meas}$}& \multicolumn{2}{c}{$F$}	&&\multicolumn{2}{c}{$F$} \\
\hline
\ozst  & 1.9579 		& 67.2&14.2 	&&&	\multicolumn{1}{c}{11.5 $\pm$ 1.10}  	\\
\ozstwo & 2.0342 	& 9.71&3.36$^{\diamond}$  	&&&	\multicolumn{1}{c}{---}  	\\
\ozso  & 2.1222 		& 91.5&6.16 	&&&	9.69 $\pm$ 0.76  		\\
\toso & 2.2481	& 7.73&3.38$^{\ddagger}$  	&&&  	\multicolumn{1}{c}{---} 	\\
\ozqo  & 2.4071 		& 46.4&14.5  	&&&  	\multicolumn{1}{c}{11.9 $\pm$ 2.95} 	\\
\ozqt  & 2.4241 		& 42.2&24.0 	&&&  	\multicolumn{1}{c}{13.3 $\pm$ 4.42} 	\\
\hline
\end{tabular}
\begin{tablenotes}
\item [$\ddagger$] Flux measured for Western H$_2$ region only.
\item [$\diamond$] Measurement confined to elliptical region
(Fig.~\ref{fig:oh231_naco_sin_wfpc2} [d]).
\end{tablenotes}
\label{tab:h2_fluxes}
\end{threeparttable}
\end{table}

Table~\ref{tab:h2_fluxes} presents the flux measurements for the central and
Northern fields, for lines with flux errors less than \simm 50 per cent. 
Accurate \tost, \ozsoo, and \ozqtwo line flux measurements are not
possible due to the 
presence of strong sky subtraction residuals at these wavelengths.
We confine the calculation of the \rto~ratio for the central field to
the area marked by the ellipse in Fig.~\ref{fig:oh231_naco_sin_wfpc2} (d), where
the \toso flux error is smallest. In this region we calculate a \rto~ratio of 8.3$\pm$1.9 prior to 
extinction correction. We do not detect any \toso flux in the Eastern H$_2$
region, instead we estimate an upper \toso limit (3$\sigma$) for the flux in this region
of 1.3$\times$10$^{-19}$ W m$^{-2}$, which in turn places a lower limit of \simm
9.5 on the \rto~ratio for this region.

\section{Analysis and Discussion}

The \rto~ratio is a useful discriminator of excitation mechanisms. 
Pure fluorescence will yield a value of $\approx$ 2 while a value $\approx$ 10
indicates the excitation of the gas is being driven by
shocks. 
However it is worth noting, these values depend on shock velocity and pre-shock gas
density, with shocked H$_2$ capable of producing values as low as
$\approx$4~\citep{1995A&A...296..789S}, while fluorescence can produce values
approaching those of shocks~\citep{1995ApJ...455..133H}.
The \rto~ratio values, given
above, suggest that shocks might be the main excitation mechanism, which agrees
with the detection of shock-excited \hco~in the centre of OH231
noted in the position-velocity diagrams of \citet{2000A&A...357..651S}.  
In order to determine the intrinsic \rto~ratio, it is
necessary to remove the effects of extinction. In the \kband, it is sometimes possible to
derive the level of extinction via the comparison of S-/Q-branch H$_2$ emission
lines~\citep[see][]{2003ApJ...592..245S}. Unfortunately, due
to poor atmospheric transmission above 2.4 $\umu$m, it
was not possible to derive a sensible estimate for extinction using the
1$\to$0~S(1) and Q(3) lines.

Extinction values of $A_{\rm K}$=3--4 (mag) for the central region are estimated from
the $(K - L^{\prime})$ colour map of OH231 \citep{1998AJ....116.1412K}.
Although, most likely an over-estimate, adjusting the Western \rto~ratio for these levels of
extinction yields, for example, a value of 11.7$\pm$2.6 ($A_{\rm K}$=4). It is clear that
any adjustment for extinction will increase the observed \rto~ratio, pushing it
further towards the shock regime.

Using a typical value of 5.3$\times$10$^{-22}$ mag cm$^{2}$ for the extinction
per unit column density of hydrogen, $A_{\rm V}/N({\rm H})$, one can estimate
the hydrogen column density implied by an $A_{\rm K}$=4;
yielding a $N({\rm H})$ for the H$_2$ regions of \simm 7.2$\times$10$^{22}$ cm$^{-2}$. 
Using a typical column length of the H$_2$ emitting regions of 1\arcsec
($\approx$2.0$\times$10$^{16}$ cm), as used in \citet{2002A&A...389..271B}, 
we estimate an average density for the H$_2$ regions of 
3.5$\times$10$^{6}$ cm$^{-3}$. This value is in good agreement with \citet{2001A&A...373..932A} 
who estimate an average central density of \simm 3.0$\times$10$^{6}$
cm$^{-3}$. The H$_2$ emission is most likely originating in the dense equatorial regions
surrounding the central star.

\subsection{Equatorial Region}
\label{sec:h2eq}

The \ozso H$_2$ line map (see Fig.~\ref{fig:oh231_naco_sin_wfpc2} [d]) shows two regions of H$_2$ oriented at a 
position angle (PA) of 113$^\circ$, from brightest
to faintest peak. 
If we assume that the distribution of the H$_2$ around the central region of
OH231 is in a disc configuration then by measuring the major and minor axis we can
estimate the
inclination angle of the equatorial disc with respect to the plane
of the sky from the H$_2$ data.
Measurements for \Rmax/\Rmin~are determined by superposing a full ellipse onto
the H$_2$ line map. We find \Rmax=
4.06\arcsec~and \Rmin= 2.76\arcsec from the centre of the ellipse, and using the relation $i =$ \asin(\Rmin/\Rmax), we
find $i=$43$^\circ$$\pm$8. 
This result is in good agreement with previously published values.

The location and orientation of this H$_2$ adds to the already complex
picture of the equatorial region of OH231. Some previously reported
structures, from smallest to largest, include: 
\begin{list}{\makelabel{-}}{\leftmargin=1em \itemsep=-0.2em}
\item an equatorial torus of SiO maser emission (\simm 2R$_{\star}$) that might lie
on the innermost edge of an expanding SO disc~\citep{2000A&A...357..651S,2002A&A...385L...1S};
\item a centrally located compact disc of circumstellar material with inner R=0.03--0.04\arcsec/40--50 AU \citep{2006ApJ...646L.123M}; 
\item an opaque flared disc with outer R=0.25\arcsec/330 AU revealed in mid-IR images~\citep{2002ApJ...574..963J};
\item a slowly expanding disc with characteristic R=0.9\arcsec/1160 AU, detected via the SO 
($J$=2$_2$$\to$1$_1$) transition \citep{2000A&A...357..651S};
\item a torus of OH maser emission with outer R=2.5\arcsec/3250 AU~\citep{2001MNRAS.322..280Z};
\item an expanding hollow cylinder of \hco~with a characteristic radius equal to the OH torus radius
~\citep{2000A&A...357..651S}.
\item a halo of scattered light at R\simm 4\arcsec/5200~AU~\citep{2003ApJ...585..482M};
\end{list}

\noindent From our observations the geometry of the H$_2$ region is unclear, however we
offer two possibilities: 
(1) A disc of H$_2$: Figure~\ref{fig:oh231_naco_sin_wfpc2} (d) shows what might be interpreted as an incomplete disc
of H$_2$, which fits with the series of concentric disc/tori structures listed
above. To understand why we observe only emission from the edges and not the
front of the disc, a comparison of the noise in the continuum at the front of the
putative disc, to the amplitude of the \ozso line peak shows both to be of the same
order. We might then attribute the `missing' H$_2$ in this region to variations in
the continuum. We would not expect to observe the back of the disc due to the
high level of extinction through the nebula. (2) A shell of H$_2$: if
the H$_2$ is situated in an axisymmetric shell 
surrounding the central star, we might explain the geometry of the H$_2$ regions
by assuming a density contrast between the poles and equator. This, combined with an increased column
depth at the edge of the shell would manifest itself as two arcs of H$_2$
emission situated equatorially~\citep[as noted by][for IRAS~19306+1407]{2005ASPC..343..282L}. This is
supported by the detection of a shell of
higher density gas and dust at the same location as the H$_2$~\citep{2003ApJ...585..482M}. 
Both scenarios offer plausible explanations for the geometry of the H$_2$
emitting region, however further observations are needed in order to favour one.

\begin{figure}
\begin{center}
\includegraphics[width=0.45\textwidth,height=0.47\textwidth,angle=0]{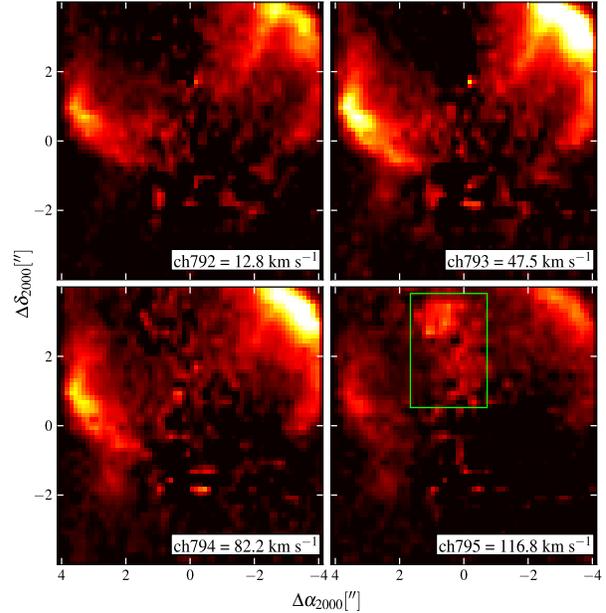}
\caption{`Channel maps' of the four central
channels of the \ozso line from the central region of OH231. Inset labels show
datacube channel and corresponding $V_{\rm LSR}$ velocity. 
Note: in ch794, the 
central region (marked by a green box) is residual from the continuum subtraction and not true H$_2$
emission.}
\label{fig:oh231_ch_maps_cen}
\end{center}
\end{figure}

We fit the \ozso line profile yielding a $V_{\rm {LSR}}$=36$\pm$17 \kms and 
FWHM=100 \kms for both Eastern and Western regions of H$_2$ emission, a value
consistent with the systemic velocity.  
Channel maps of the \ozso line (Fig.~\ref{fig:oh231_ch_maps_cen}) show no
significant change in the distribution of the H$_2$.

In an attempt to explain the lack of reported H$_2$ in this object, we note 
two previous studies: (1) \citet{1998ApJ...509..728W} give a 3$\sigma$ upper limit of 10$^{-5}$
ergs cm$^{-2}$ s$^{-1}$ ster$^{-1}$ (1$\sigma$ limit =6.6$\times$10$^{-18}$ W m$^{-2}$) 
for the surface brightness of the \ozso line towards OH231. This places their
measurement limit close to the \ozso line strength given in
Table~\ref{tab:h2_fluxes}, implying that the \ozso line in their observations
would be difficult to distingush from noise, or possibly that their 
slit position, aligned East-West across inferred central star position, did not
include the H$_2$ sites; 
(2) \citet{2006ApJ...646L.123M}, using the {\sc isaac} instrument on VLT, do not
report any detection of H$_2$. However, this can be explained due to the orientation of
the slit, aligned from South-West to North-East along the major axis (private
comm.), with a slit-width of 0.8\arcsec, i.e., the central H$_2$ emission site was not covered.

\subsection{Northern Region}
\label{sec:h2Nor}

Figure~\ref{fig:oh231_naco_sin_wfpc2} (c) shows the line map for the
\ozso transition in the Northern region. Most notable are the two knots
of H$_2$, labeled A and B for the top and bottom knot, respectively. 
The \halpha~emission contours
are superposed on the H$_2$ line map. The peak intensity of the H$_2$ emission knots are slightly offset
from the two \halpha~emission knots (Fig.~\ref{fig:oh231_naco_sin_wfpc2} [c]), however this small offset can be accounted
for by the motion of the outflow \citep[e.g.$V$\simm 150 \kms, from][]{2002A&A...389..271B} during the 10 years between both sets of
observations. It is most likely that the optical and NIR line emission arise from the same shock event. 
Weaker \ozso emission
is noted tracing the \halpha~edge of the bipolar outflow in the lower portion of
the \ozso line map.  
The NACO 2.12~$\umu$m image does not show any trace of H$_2$ emission in this
region (see contours in Fig.~\ref{fig:oh231_naco_sin_wfpc2} [b]). 

\begin{figure}
\begin{center}
\includegraphics[width=0.45\textwidth,height=0.47\textwidth,angle=0]{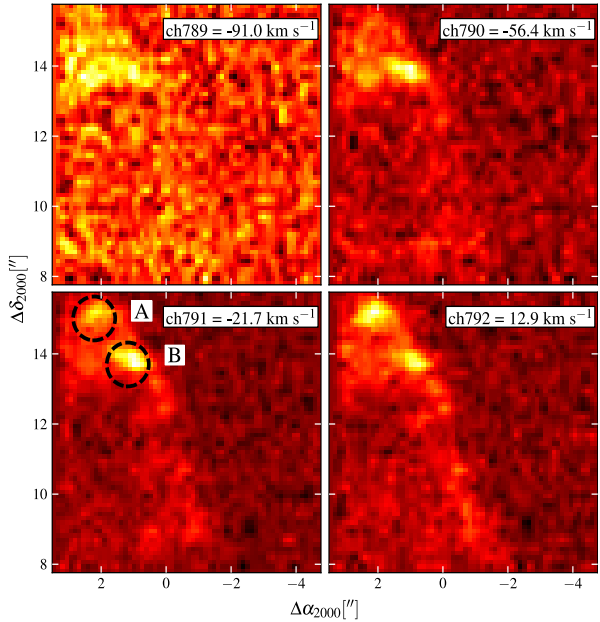}
\caption{\ozso line `channel maps' of the four central channels extracted 
from the Northern region of OH231. Knot A and B are regions of most intense
H$_2$ emission. Labels as in Fig.\ref{fig:oh231_ch_maps_cen}.
}
\label{fig:oh231_ch_maps}
\end{center}
\end{figure}

The \ozso line is spread over \simm six spectral pixels, the four central channels are
presented in Fig.~\ref{fig:oh231_ch_maps}. Examination of the
line profile and channel maps allows us to probe the  
kinematics of the H$_2$ in this region, revealing two main features:
\begin{inparaenum}[1)]
\item{knot B which is persistent in all channel maps }
\item{the reduction of peak H$_2$ intensity in knot A at velocities $\leq$~-56
\kms.}
\end{inparaenum}
Both of these H$_2$ structures lie in the diffuse extended region
\citep[labeled B$_1$ in Fig.~4 of ][]{2002A&A...389..271B} 
perpendicular to the axial flow, with a quoted \halpha~velocity of 150
\kms~which is in good agreement with not only our deprojected H$_2$
velocities 
($V_{\rm {H_2}}$\simm 110 \kms)
but also with 
\hco~velocities \citep{2000A&A...357..651S}.

In the case of knot A, the spectral line is strongly peaked in a single
channel, while in knot B the spectral line peak is spread over two channels.
Fitting the line profiles of knots A and B, yields LSR velocities of
$V_{\rm {LSR}}$=-8 \kms, and -30 \kms, with deprojected velocities of -75 \kms and -110 \kms, respectively. 
This might indicate that further from the \halpha~bow
apex (Fig.~\ref{fig:oh231_naco_sin_wfpc2} [a]), we are starting to see emission originating from the front and back of
knot B, while knot A displays a narrower range of velocities, i.e., a single peak in its spectral profile.
It is worth noting that due to the slit-length of {\sc isaac}
(120\arcsec) coupled with the null H$_2$ detection discussed in \S\ref{sec:h2eq}, the H$_2$ in
the Northern region might be confined to the wings of the bow shock.

The existence of fast moving shocked H$_2$ has
previously been noted in other objects,  
for example, \citet{2003ApJ...586L..87C} detect high velocity H$_2$
\simm 220--340 \kms~(dependent on adopted inclination angle) in the outflows of AFGL~618. It is currently unknown exactly how
shocked H$_2$ can be travelling at this speed without complete dissociation.
Further high resolution mapping of the H$_2$ is necessary in order to 
resolve the shock surfaces, more accurately measure the H$_2$ kinematics, and apply shock models to this region.

\section{Conclusions}
\label{sec:discuss}

We have presented VLT/{\sc sinfoni} integral field observations of OH231,
revealing the presence of several ro-vibrational H$_2$ lines. 
The main conclusions are:
\begin{list}{\makelabel{-}}{\itemsep=-0.2em \leftmargin=1em}
\item The discovery of H$_2$ emission near the centre of OH231, possibly
located at the edge of an axisymmetric shell or an incomplete disc.
\item A \rto~value of 8.3$\pm$1.9 is found for the equatorial H$_2$, suggesting
a collisional excitation mechanism. 
\item Our observations of the central shell/disc of H$_2$ show no velocity structure. 
However, higher S/N and/or velocity resolution data are needed to accurately
probe the kinematics in this region. 
\item We detect fast-moving H$_2$ (\simm 110 \kms, along the bipolar axis) via the \ozso transition along the
North-Western tip of the nebula, a region where a strong \halpha~bow shock exists. 
Due to the small FOV of our observations, the full extent of this H$_2$ is unknown. 
\end{list}

\section{Acknowledgments}
\label{sec:Acknow}
This research is funded by UH studentship and based on observations made 
with ESO Telescopes at the Paranal Observatory under programme ID 084.D-0487(A)
and 
072.D-0766(A). We thank Mikako Matsuura for providing {\sc isaac} observation information.
This research used the HLA (ID 8326) facilities of the STScI, the ST-ECF and 
the CADC with the support of the following granting agencies: NASA/NSF, ESA,
NRC, CSA.

\bibliographystyle{mn2e_orig}

\end{document}